\documentclass[prb, aps, twocolumn, groupedaddress, superscriptaddress]{revtex4}

\usepackage{slashed}
\usepackage{graphicx}
\usepackage{subfigure}
\usepackage[usenames, dvipsnames]{color}
\usepackage{graphics}
\usepackage{hyperref}
\usepackage{bm}
\usepackage{amsfonts}
\usepackage{lipsum}
\hypersetup{backref,
colorlinks=true,
linkcolor=blue,
linktoc=page,
citecolor=blue}

\begin{document}

\title{Slow sound in matter-wave dark soliton gases}

\author{Muzzamal I. Shaukat}
\affiliation{ Instituto Superior T\'ecnico, University of Lisbon and Instituto de Telecomunica\c{c}\~{o}es, Torre Norte, Av. Rovisco Pais 1,
Lisbon, Portugal}
\affiliation{CeFEMA, Instituto Superior T\'ecnico, Lisbon, Portugal}
\affiliation{University of Engineering and Technology, Lahore (RCET Campus), Pakistan}
\email{muzzamalshaukat@gmail.com}
\author{Eduardo V. Castro}
\affiliation{CeFEMA, Instituto Superior T\'ecnico, Universidade de Lisboa, Lisboa, Portugal}
\affiliation{Centro de F\'isica das Universidades do Minho e Porto,
Departamento de F\'isica e Astronomia, Faculdade de Ci\'encias,
Universidade do Porto, Porto, Portugal}
\author{Hugo Ter\c{c}as}
\affiliation{Instituto de Plasmas e Fus\~ao Nuclear, Instituto Superior T\'ecnico, Lisboa, Portugal}
\email{hugo.tercas@tecnico.ulisboa.pt}


\begin{abstract}

We demonstrate the possibility of drastically reducing the velocity of phonons in quasi one-dimensional Bose-Einstein condensates. Our scheme consists of a dilute dark-soliton ``gas" that provide the trapping for the impurities that surround the condensate. We tune the interaction between the impurities and the condensate particles in such a way that the dark solitons result in an array of {\it qutrits} (three-level structures). We compute the phonon-soliton coupling and investigate the decay rates of these three-level qutrits inside the condensate. As such, we are able to reproduce the phenomenon of acoustic transparency based purely on matter wave phononics, in analogy with the electric induced transparency (EIT) effect in quantum optics. Thanks to the unique properties of transmission and dispersion of dark solitons, we show that the speed of an acoustic pulse can be brought down to $\sim 5$ $\mu$m/s, $\sim 10^3$ times lower than the condensate sound speed. This is a record value that greatly underdoes most of the reported studies for phononic platforms. We believe the present work could pave the stage for a new generation of ``stopped-sound" based quantum information protocols.  

\end{abstract}

\maketitle

\section{Introduction}

Electromagnetically induced transparency (EIT) \cite{Harris1990} is a quantum interference effect in which the absorption of a weak probe laser, interacting resonantly with an atomic transition, is reduced in the presence of a coupling laser. EIT plays a crucial role in the optical controll of slow light \cite{Hau1999} and optical storage \cite{Phillips2001}, having been extensively investigated in $\Lambda$-, V- and cascade-type three-level systems \cite{Abi2010,Anisimov2011}. This fascinating effect has been experimentally observed in both atoms \cite{Boller1990} and semiconductor quantum wells \cite{Serapiglia2000}. A major problem in the initial studies of EIT in atomic vapors has to do with the thermal spectral broadening \cite{Cornell02, Ketterle02}, smearing out the EIT window. In order to mitigate this issue, researchers have made use of coherent Bose-Einstein condensates (BECs) \cite{Vadeiko2005, Ri2007, Ahufinger2002}. The association of EIT with light-matter coupling can be used to prepare and detect coherent many-body phenomena in ultra-cold quantum gases \cite{Ruostekoski1999}.\par

Soon after the engineering of photonic crystal structures, the attention has been drawn to the propagation of acoustic waves in periodic media \cite{Lheurette2013, Craster2013}. Many intriguing phenomena, such as the analogue of EIT \cite{Liu2009, Fleischhauer2005} and Fano resonances \cite{ Lukyanchuk2010, Khscianikaev2011} have been envisaged in the context of acoustics as well \cite{santillan2011, Amin2015}. For example, an isotropic metamaterial consisting of grooves on a square bar traps acoustic pulses due to a strong modulation of wave group velocity \cite{Zhu13}; slowing down the speed of sound in sonic crystal waveguides has also been achieved, with a reported group velocity of $26.7$ m/s \cite{Cicek12}. Soliton propagation and soliton-soliton interaction in EIT media has been studied by Wadati et al. \cite {Wadati2009}, and 
the formation of solitons via dark-state polaritons has been proposed \cite{Xiong2004}.\par

Recently, we have shown that a dark-soliton (DS) qubit in a quasi one-dimensional (1D) BEC is an appealing candidate to store quantum information, thanks to its appreciably long lifetimes ($\sim 0.01-1$s) \cite{Muzzamal2017}. Moreover, we explored the creation of quantum correlation between DS qubits displaced at appreciably large distances (a few micrometer) \cite{Muzzamal2018a, Muzzamal2018b, Muzzamal2019}. Dark-soliton qubits thus offer an appealing alternative to quantum optics in solid-state platforms, where information processing involves only phononic degrees of freedom: the quantum excitations on top of the BEC state.\par 

\begin{figure}[t!]
\includegraphics[width=0.4\textwidth]{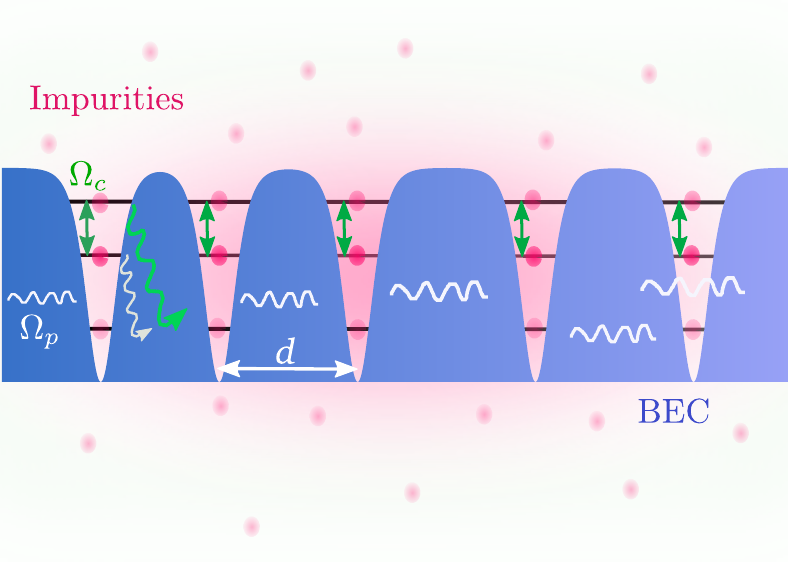}
\caption{(color online) Schematic representation of dark-soliton qutrit array immersed in a BEC. The impurities (free particles) are trapped inside dark soliton potential. The wiggly lines represent the quantum fluctuations on top of the condensate (phonons).}
\label{fig_model}
\end{figure}

In this paper, we propose to make use of dark solitons to achieve a phenomenon with EIT-like characteristics, the {\it acoustic transparency} (AT). The active medium is composed of a set of dark-soliton {\it qutrits}, i.e. three-level objects comprising an impurity trapped at the interior of a dark-soliton potential. Quantum fluctuations are provided by the BEC acoustic (Bogoliubov) modes, or simply phonons (see Fig. \ref{fig_model} for a schematic representation). We start by recalling the conditions under which that qutrit is achievable. Then, the qutrit array is shown to be an open quantum system, where the reservoir is composed by the BEC phonons \cite{stringari_book}. We compute the linewidth of each of the qutrit transitions by treating the qutrit-phonon interaction within the Born-Markov approximation. We conclude by computing the dispersion relation of a weak envelope of sound waves and show that its group velocity can be drastically reduced to $\sim 0.06$ mm/s, to the best of our knowledge a record value ever reported in acoustics. Our study represents an advance in the direction of `slow-sound' schemes and the results have potential applications in phononic information processing. \par
The paper is organized as follows: In sec. II, we start with the set of coupled Gross-Pitaevskii and Schr\"odinger equations, to study the properties of dark solitons in quasi-1D BEC, imprinted in a dilute set of impurities. The coupling between DSs and phonons is computed in Sec. III,  followed by a discussion on 
spontaneous decay of three level system in Sec. IV. The concept of slow sound due to quantum interference phenomenon is described in Sec. V.
We conclude the practical implications of the present scheme in sec. VI.


\section{Dark-soliton qutrits} 

We starting by considering a dark soliton in a quasi 1D BEC, with the later being surrounded by a dilute gas of impurities (see Fig. \ref{fig_model}). The DS plays the role of a potential for the impurities (considered to be free particles) and the phonons act like a quantum (zero-temperature) reservoir. The solitons and the impurities can be treated at the mean field level, being respectively governed by the Gross-Pitaevskii and the Schr\"odinger equations,
\begin{eqnarray}
i\hbar \frac{\partial \psi _{1}}{\partial t}&=&-\frac{\hbar ^{2}}{2m_1}\frac{
\partial^{2} \psi _{1}}{\partial x^{2}}+g_{11}\left\vert \psi _{1}\right\vert ^{2}\psi
_{1}+g_{12}\left\vert \psi _{2}\right\vert ^{2}\psi _{1}, \nonumber \\ 
i\hbar \frac{\partial \psi _{2}}{\partial t}&=&-\frac{\hbar ^{2}}{2m_2}\frac{
\partial^{2} \psi _{2}}{\partial x^{2}}+g_{21}\left\vert \psi_{\rm sol}\right\vert ^{2}\psi _{2}.  \label{gp2}
\end{eqnarray}
Here, $g_{11}$ represents the BEC inter-particle interaction strength, $g_{12}=g_{21}$ is the BEC-impurity coupling constant (Appendix-\ref{Trapping impurities with dark solitons}), while $m_1$ and $m_2$ denote the BEC particle and impurity masses, respectively. To distinguish the weakly interacting quasi 1D regime from strongly interacting Tonks-Girandeau gas \cite{javed2016}, the dimensionless quantity $\alpha=2a_s l_z/l_r^2\ll1$ is $\sim 0.06$ for BEC and  $\sim 0.07$ for impurity particles  in case of dark solitons. Here, $l_z$ ($l_r$) is the longitudinal (transverse) size and $a_s \sim 0-37$nm is the $^{85}{\rm Rb}$ $s$-wave scattering length \cite{Roberts2000}.

The singular nonlinear solution corresponding to the soliton profile is \cite{zakharov72, huang} 
\begin{equation}
\psi_{\rm sol}(x)=\sqrt{n_{0}}\tanh \left( \frac{x}{\xi} \right),
\end{equation}
 where $n_{0}$ denotes the BEC linear density and $\xi =\hbar /\sqrt{m_1n_{0}g_{11}}$ is the healing length. The latter lies in the range $(0.7-1.0)$ $\mu$m in a typical 1D BECs, for which the condensate is homogeneous along a trap of size $L \sim 70$ $\mu$m \cite{Gaunt}). More recent experiments leads eventual trap inhomogeneities to be much less critical by providing much larger traps, $L\sim 100$ $\mu$m \cite{schmiedmayer2010}. The time-independent version of the impurity equation in (\ref{gp2}) reads
\begin{equation}
E'\psi _{2}=-\frac{\hbar ^{2}}{2m_2}\frac{%
\partial^{2} \psi _{2}}{\partial x^{2}}-g_{21}n_{0}{\rm sech}^{2}\left( \frac{x}{\xi }\right) \psi _{2},
\label{eq_reflectionless1}
\end{equation}
where $E'=E - n_{0} g_{21}$ \cite{note1}. To find the analytical solution of Eq. (\ref{eq_reflectionless1}), the potential is casted in the P\"oschl-Teller form $V(x)=-\hbar ^{2}\nu(1+\nu)  {\rm sech}^{2} \left(x/\xi\right)/2m_2 \xi^{2}$, with $2\nu=-1+\sqrt{1+4g_{21}m_2/g_{11}m_1}$ and the energy spectrum $E'_{n}= -\hbar ^{2}\left(\nu-n\right)^{2}   /2m_2 \xi^{2}$, where $n$ is an integer \cite{john07}. The number of bound states created by the DS is $n_{\rm bound}=\lfloor \nu+1+\sqrt{\nu(1+\nu)}\rfloor$, where the symbol $\lfloor \cdot\rfloor$ denotes the integer part. As such, for a DS to contain exactly three bound states (i.e. the condition for the qutrit to exist), the parameter $\nu$ must lie in the range
\begin{eqnarray}
\frac{4}{5}\leq \nu < \frac{9}{7}.  \label{condition}
\end{eqnarray}
At $\nu \geq 9/7$, the number of bound states increases. However, the effect of the impurity on the profile of the soliton itself becomes more important, and therefore special care must be taken in the choices of the mass ration $m_2/m_1$. In our numerical calculations below, we choose $^{85}$Rb BEC solitons trapping $^{134}$Cs impurities (Appendix-\ref{Trapping impurities with dark solitons}). However, other choices are possible and our analysis remains general. \par
\begin{figure}[t!]
\includegraphics[width=0.241\textwidth]{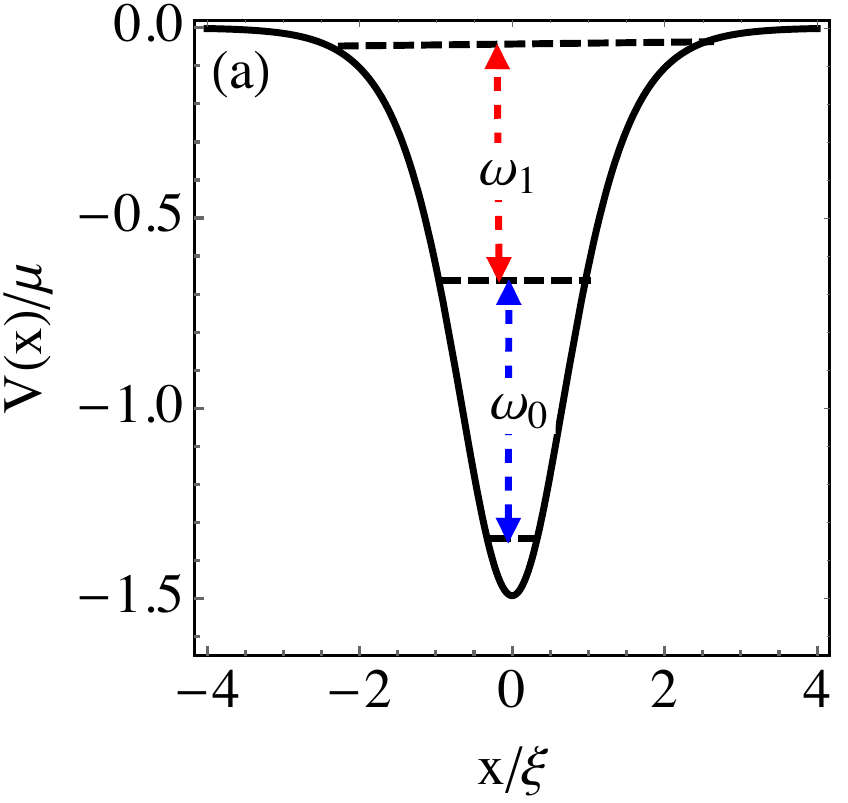}
\includegraphics[width=0.234\textwidth]{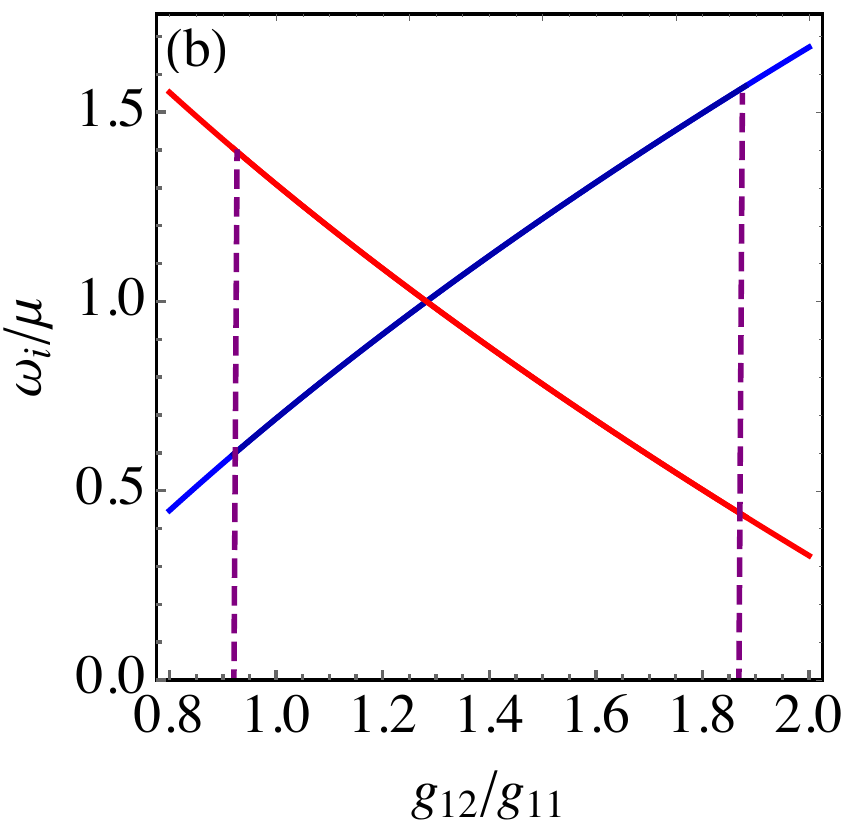}
\caption{(color online) Panel (a) illustrates the impurity states and the respective transition frequencies $\omega_i$ $(i=0,1)$ in the dark soliton potential. Panel (b) shows the dependence of the transition frequencies on the BEC-impurity coupling $g_{12}$. The vertical dashed lines corresponds to the range $4/5 \leq \nu  < 9/7$ defined for the qutrit. For definitness, we have used the $m_2=1.56 m_1$, corresponding to a $^{134}$Cs impurity loaded in a $^{85}$Rb BEC dark soliton.}\par 
\label{fig_frequency}
\end{figure}


\section{Quantum fluctuations} 

The total BEC quantum field includes the DS wave function and quantum fluctuations, $\psi_1(x)=\psi_{\rm sol}(x)+\delta \psi(x)$, where $\delta \psi(x)=\sum_k \left(u_k(x) b_k +v^{*}_k(x)b^{\dagger}_k \right)$ and $b_k$ are the bosonic operators verifying the commutation relation $[b_{k},b^{\dagger}_{q}]=\delta_{k,q}$. The amplitudes $u_k(x)$ and $v_k(x)$ satisfy the normalization condition $\vert u_k(x)\vert ^2 -\vert v_k(x)\vert ^2=1$ and are explicitly given in Appendix-\ref{Soliton-phonon Hamiltonian}. The total Hamiltonian then reads $H=H_{\rm q}+H_{\rm p}+H_{\rm int}$, where $H_{\rm q}=\hbar \omega _{1} \left(\vert e_{2}\rangle  \langle e_{2}\vert-\vert e_{1}\rangle\langle e_{1} \vert\right) +\hbar \omega _{0} \left(\vert e_{1}\rangle \langle e_{1}\vert -\vert g\rangle \langle g\vert \right)$ is the qutrit Hamiltonian, with $\omega _{1}=\hbar(2\nu -3)/(2m\xi ^{2})$ and $\omega _{0}=\hbar(2\nu -1)/(2m\xi ^{2})$ are the gap energies for  $\left\vert e_{2}\right\rangle \leftrightarrow \left\vert e_{1}\right\rangle$ and $\left\vert e_{1} \right\rangle \leftrightarrow \left\vert g \right\rangle$ transitions,  respectively. The term $H_{\rm p}=\sum_k \epsilon _{k}b_{k}^{\dagger} b _{k}$ represents the phonon (reservoir) Hamiltonian, where $\epsilon _{k}=\mu \xi \sqrt{k^{2}(\xi^{2}k^{2}+2)}$ is the Bogoliubov spectrum with chemical potential $\mu=g_{11}n_{0}$. The interaction Hamiltonian is given by
\begin{equation}
H_{\rm int}=g_{12}\int dx \psi_{2} ^{\dag }\psi_1 ^{\dag }\psi_1 \psi_2,
\label{Int. Ham.}
\end{equation}
where $\psi_2(x)=\sum_{l=0}^2 \varphi_{l}(x) a_{l}$ describes the impurity field spanned in terms of the bosonic operators $a_{l}$, with $\varphi_0(x)=A_0{\rm sech}  ^{\alpha}(x/\xi)$, $\varphi_1(x)=2 A_1\tanh  (x /\xi) \varphi_0(x)$ and $ \varphi_2(x)= \sqrt{2}A_2\left(1-(1+3\alpha)\tanh^{2} (x/\xi) \right)\varphi_0(x)$, where $A_j (j=0,1,2)$ are the normalization constants and $\alpha=\sqrt{2g_{12}m_2/(g_{11}m_1)}$ (see Appendix-\ref{Soliton-phonon Hamiltonian}). Using the rotating wave approximation (RWA), the first order perturbed Hamiltonian can be written as
\begin{eqnarray}
H_{\rm int}^{(1)} &=&\sum_{k}\left(g_{0}^{k}\sigma^{+} _{0} +g_{1}^{k}\sigma^{+} _{1}\right) b_{k}+ \left( g_{0}^{k*}\sigma^{-} _{0} + g_{1}^{k*}\sigma^{-} _{1}\right)b_{k}^{\dag }, \nonumber 
\label{eq_ham_int1}
\end{eqnarray}
where  $\sigma^{+} _{0,1}=a^{\dagger}_{e_1,e_2} a_{g,e_1} $,  $\sigma^{-} _{0,1}=a^{\dagger}_{g,e_1} a_{e_1,e_2} $, while the coupling constants $g_{ll'}^{k}=g_{i}^{k}$($i=0,1$) are explicitly given in Appendix-\ref{Soliton-phonon Hamiltonian}.
%
In our RWA calculation, the counter-rotating terms proportional to $b_{k}\sigma_{i} ^{-}$ and $b^{\dagger}_{k}\sigma_{i} ^{+}$ are dropped. The accuracy of such an approximation can be verified  a \textit{posteriori}, provided that the emission rates $\gamma_{0}$ and $\gamma_{1}$ are much smaller than the qutrit transition frequencies $\omega_{0}$ and $\omega_{1}$, respectively.
\begin{figure}[t!]
\includegraphics[width=0.4\textwidth]{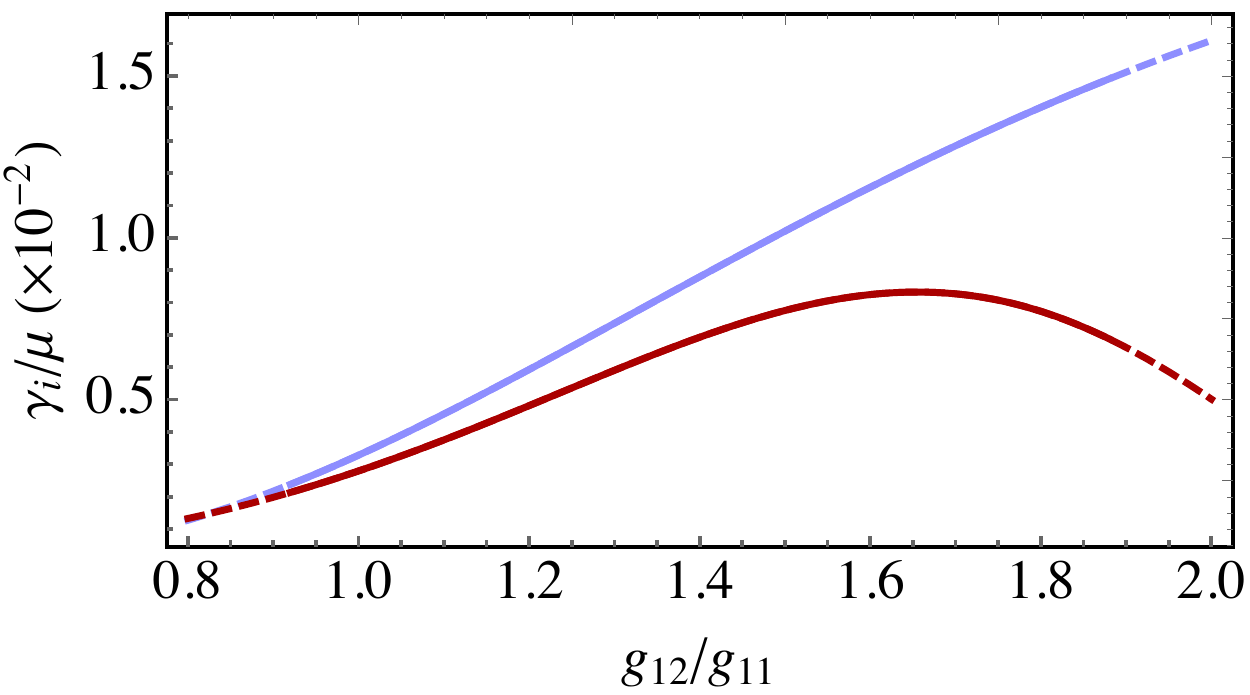}
\caption{(color online) Dependence of the decay rates $\gamma_{0}$ (blue line) and $\gamma_{1}$ (red line) on the BEC-impurity coupling. The solid lines correspond to the range $\frac{4}{5} \leq \nu  <\frac{9}{7}$ that defines the qutrit. We have used $m_2=1.56 m_1$, as in Fig. \ref{fig_states}.}
\label{fig_decay}
\end{figure}


\section{Wigner-Weisskopf theory of spontaneous decay}

We  employ  the  Wigner-Weisskopf  theory to find the spontaneous decay rate of the states, by neglecting  the  effect  of  temperature  and  other  external
perturbations \cite{scully_book}. In this regard, the qutrit is assumed to be initially at the excited state $\left\vert e_{2}\right\rangle$ and the phonons to be in the vacuum state $\left\vert 0 \right\rangle$. Under such conditions, the wave function of total system (qutrit + phonons) can be described as
\begin{eqnarray}
\left\vert \Phi(t) \right\rangle &=& a(t) \left\vert e_{2},0 \right\rangle + \sum_{k} b_{k}(t) \left\vert e_{1},1_{k} \right\rangle \nonumber \\ 
&+& \sum_{k,p} b_{k,p}(t) \left\vert g,1_{k},1_{p} \right\rangle,
\end{eqnarray}
where $a(t)$ is the probability amplitude of the excited state $\left\vert e_{2}\right\rangle$. The qutrit decays to the state $\left\vert e_{1}\right\rangle$ with probability amplitude $b_{k}(t)$ by emitting a phonon of wavevector $k$ and frequency $\omega_k$. Subsequently, the qutrit de-excites to the ground state $\left\vert g \right\rangle$ via the emission of a phonon of momentum $p$, frequency $\omega_p$ and probability amplitude $b_{k,p}(t) $. In the interaction picture, these coefficients can be written as (Appendix-\ref{Wigner-Weisskopf theory of spontaneous decay}),
\begin{eqnarray}
a(t)&=& e^{-\gamma_{1}t/2}, \nonumber \\
b_{k}(t)&=&-ig_{0}^{k}\frac{\left[e^{i(\omega_{k}-\omega_{1})t-\gamma_{1}t/2}-e^{-\gamma_{0}t/2}\right]}{i(\omega_{k}-\omega_{1})-\frac{\gamma_{1}-\gamma_{0}}{2}}, \nonumber \\
b_{k,p}(t)&=&\frac{g_{0}^{k}g_{1}^{k}}{i(\omega_{k}-\omega_{1})-\frac{\gamma_{1}-\gamma_{0}}{2}}
\left[ \frac{e^{i(\omega_{p}-\omega_{0})t-\gamma_{0}t/2}- 1 }{i(\omega_{p}-\omega_{0})-\frac{\gamma_{0}}{2}}  \right. \nonumber \\
&&\left. +  \frac{1-e^{i(\omega_{k}+\omega_{p}-\omega_{0}-\omega_1)t-\gamma_{1}t/2}}{i(\omega_{k}+\omega_{p}-\omega_{0}-\omega_1)-\frac{\gamma_{1}}{2}} \right],  \label{elements}
\end{eqnarray}
where $\gamma_{i}$ $(i=0,1)$ is the $i$th state decay rate
\begin{eqnarray}
\gamma_i &=&\frac{L}{\sqrt{2}\hbar \xi}\int ~ d\omega_k  \frac{\sqrt{1+ \eta_{i}}}{\eta_{i}}\vert g_i^k\vert^2 \delta(\omega_k-\omega_i), 
\label{eq_gamma}
\end{eqnarray}
where $\eta_{i}=\sqrt{\mu^{2}+\hbar ^{2}\omega _{i}^{2}}$. The validity of both the RWA and the Born-Markov approximations is illustrated in Figs. \ref{fig_frequency} and \ref{fig_decay}, where it is depicted that the decay rates of both transitions are much smaller than the respective transition frequencies.
The soliton retains its shape has confirmed by Javed et. al. \cite{javed2016}, while 
investigating the quasi-1D model of $^{133}{\rm Cs}$ impurities in the center of a trapped $^{87}{\rm Rb}$ BEC. Moreover, the occurrence of impurity condensation on the bottom of the soliton, due to a sufficiently high concentration of impurities, leads to the breakdown of single particle assumption and spurious energy shift. This can be avoided if fermionic impurities are used instead \cite{Hansen2011}.
It is pertinent to mention here two experimental considerations. First, notice that Feshbach resonances can be used to tune the value of $g_{12}$, allowing for an additional control of the rates $\gamma_i$. Second, dark-soliton quantum diffusion may be the only immediate limitation to the performance of the qutrits \cite{Dziarmaga04}, a feature that has been theoretically predicted but yet not experimentally validated. In any case, quantum evaporation is expected if important trap anisotropies are present, a limitation that we can overcome with the help of box-like or ring potentials \cite{Gaunt}. This is exactly the situation we will consider in the numerical calculations below.


\section{Acoustic Bloch equations}

To produce the interference effect necessary for the AT, we consider the situation where the qutrit is externally driven by two (probe and control) acoustic fields by following the scheme of Raman lasers, described in Ref.'s \cite{Sabin2014, Recati2005, Compagno_2017}. The DS qutrit states are coupled to the phonons with a Raman laser driving the transition $\left\vert g \right\rangle  \leftrightarrow \left\vert e_{1} \right\rangle $ with frequency $\omega_{p}$ and detuning $\Delta_{p}=\omega_{p}-\omega_{0}$ to interact with BEC.  Simultaneously, a Raman laser field of frequency $\omega_{c}$ and detuning $\Delta_{c}=\omega_{c}-\omega_{1}$ couple the states $\left\vert e_{1} \right\rangle  $ and  $\left\vert e_{2} \right\rangle$. Therefore, the qutrit driving can be described, within the RWA approximation, by the following Hamiltonian
\begin{eqnarray}
H_{\rm drive} &=&\frac{\hbar}{2}\left(\Omega_{p}\left\vert e_{1} \right\rangle \left\langle g \right\vert+\Omega_{c}\left\vert e_{2} \right\rangle \left\langle e_{1} \right\vert -2 \Delta_{p}\left\vert e_{1} \right\rangle \left\langle e_{1} \right\vert \right. \nonumber \\
&&\left. - 2\delta \left\vert e_{2} \right\rangle \left\langle e_{2} \right\vert\right)+ {\rm H.c.},
\end{eqnarray}
where $\delta = \Delta_{p}+\Delta_{c}$ and $\Omega_{p,c}$ denote the Rabi frequency of the probe and control fields, respectively. We obtain the solution for the density matrix $\rho$ by solving the master equation
\begin{eqnarray}
\dot{\rho_{q}}(t)  = -\frac{i}{\hbar} \left[H_{q},\rho_{q}(t)\right] + \sum_{i=0}^{1} \gamma_{i} \mathcal{L}_{i}[\rho]
 \label{master eq.},
\end{eqnarray}
with $\rho_{ij}=\rho^{*}_{ji}$ and the Lindblad operator $\mathcal{L}[\rho]=\left[\sigma_{-}\rho_{q}(t)\sigma_{+}-\frac{1}{2} \lbrace \sigma_{+}\sigma_{-},\rho_{q}(t) \rbrace \right]$. In the limit of the weak-probe approximation, $\Omega_p\ll \Omega_c$, the steady state-coherences are given by
\begin{eqnarray}
\rho_{21}&=&\frac{i\Omega_{p}} { \left(\gamma_{0}-2i\Delta_{p}\right)+\frac{\Omega_{c}^{2}}{\gamma_{1}-2i\delta}},  \nonumber \\
\rho_{31}&=&\frac{-i\Omega_{c}} {\left(\gamma_{1}-2i \delta  \right)} \rho_{21}.
\end{eqnarray}
In what follows, we consider a set of solitons, i.e. a soliton gas \cite{tercas}, of density $N=1/d$, with $d$ denoting the average distance between the solitons. If the solitons are well-separated, $d\ll \xi$, we can assume the qutrits to be independent. This is not usually the case in one-dimensional systems, unless in the especial comensurability situation, as a consequence of the infinite-range (sinusoidal) character of the collective decay rate \cite{Gonzalez2011,ramos_2014}. Fortunately in our case, because the solitons locally deplete the condensate density, the collective scattering rate vanishes at distances largely exceeding the healing length, $d \gg \xi$ \citep{Muzzamal2018a,Muzzamal2018b}. As such, we can determine the long wavelength behavior, $k d\ll 1$, of the probe field envelope. Using the Heisenberg's relation $i\hbar \partial \left(\delta \Psi\right)/\partial t=\left[\hat{H},\delta\Psi\right]$ and the fluctuating field $\delta\Psi= \phi b_q e^{iqx}+\psi^* b_q^{\dagger} e^{-iqx}$, where $\phi$ and $\psi$ are the Bogoliubov coefficients, we obtained the propagating equation (Appendix-\ref{Heisenberg's equation and sound propagation})
\begin{eqnarray}
\frac{\partial \Omega_p}{\partial t}+ \frac{\omega_{q}}{q}\frac{\partial \Omega_p}{\partial x}=-\frac{i}{2\hbar^{2}}( g^{k_{\rm res}}_0)^2\rho_{12}, \label{Propagating equation}
\label{dispersion}
\end{eqnarray}
where $\Omega_p=N\xi g_{0}^{k_{\rm res}}  \vert \delta \Psi\vert/\hbar$ and $k_{\rm res}=0.9/\xi$ is the resonant wavevector. By ignoring the time derivative from Eq. (\ref{Propagating equation}) (time-independent fluctuating field) and comparing it with $\partial_z\delta\Psi=i k\chi \delta\Psi/2$ \cite{Lambropoulos07}, we express the soliton-gas susceptibility as
\begin{eqnarray}
\displaystyle{\chi = - \frac{ i N \xi  ( g_{0}^{k_{\rm res}})^{2}}{\hbar \epsilon_{k} \left[ \left(\gamma_{0}-2i\Delta_{p}\right)+\frac{\Omega_{c}^{2}}{\gamma_{1}-2i\delta}\right]}},
\end{eqnarray}
where we have replaced $\rho_{12}$ by its mean value in the soliton gas, $$\langle\rho_{12}\rangle=\rho_{12}^{\rm gas}=N\xi\rho_{12} ,$$ containing the information about the number of solitons per unit length, $N\xi$. The acoustic response of the envelope can be determined by the refractive index $n = \sqrt{1+\chi}$.
\begin{figure}[t!]
\includegraphics[width=0.236\textwidth]{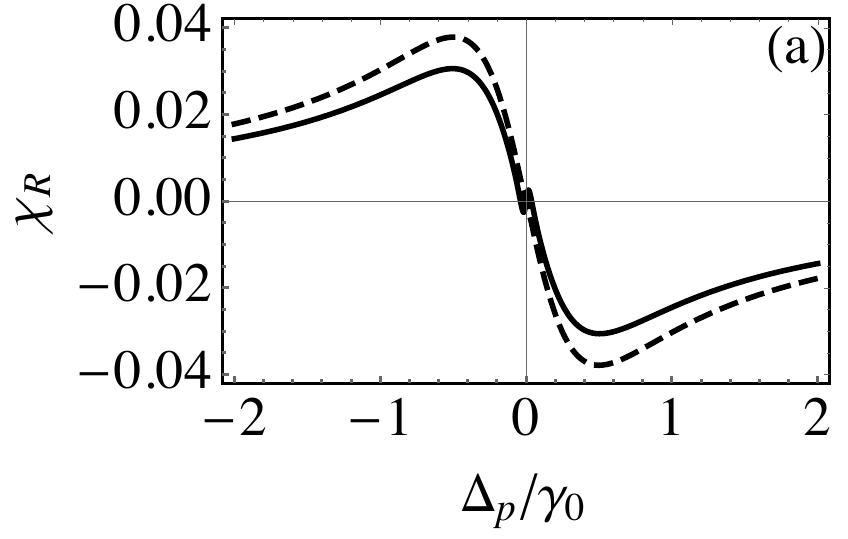}
\includegraphics[width=0.236\textwidth]{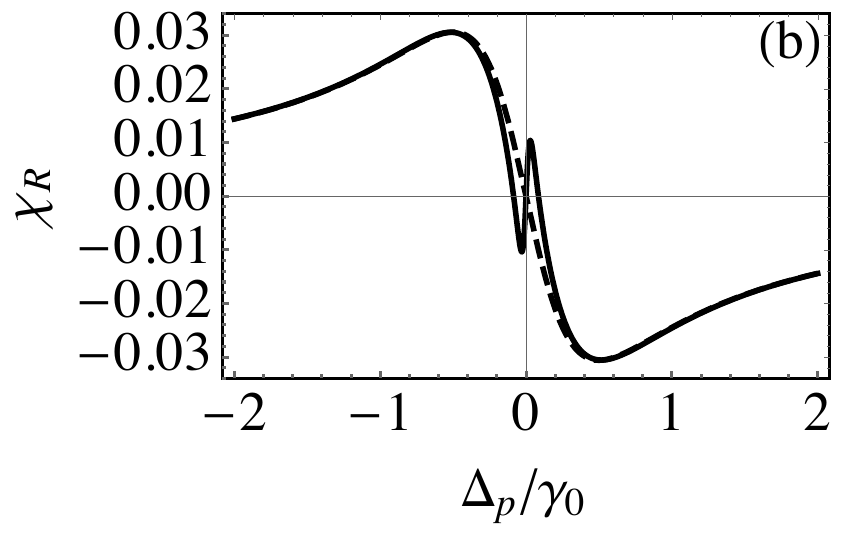}
\includegraphics[width=0.236\textwidth]{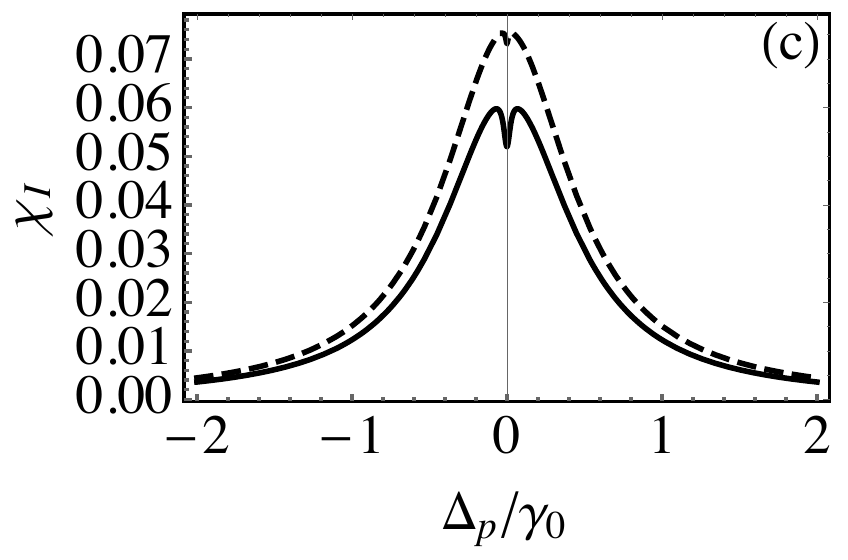}
\includegraphics[width=0.236\textwidth]{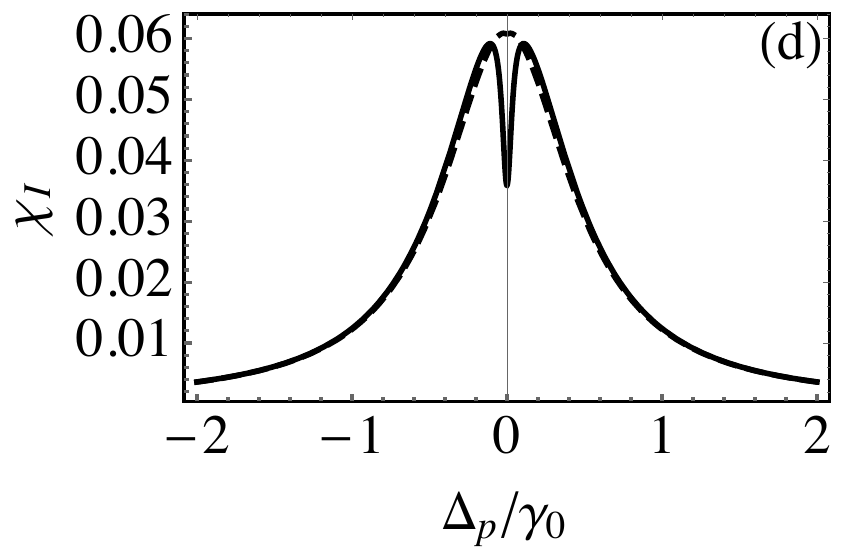}
\caption{(color online) Acoustic susceptibility $\chi$ dependence on the probe detuning $\Delta_p$. Panel (a) depicts the dispersion and panel (c) the absorption spectra calculated for $g_{12}=1.1g_{11}$ (dashed line) and $g_{12}=1.85g_{11}$ (solid line), with $\Omega_c >\gamma_1$. Panel (b) shows the dispersion and panel (d) the absorption for $\Omega_c=0.2\gamma_0$ (dashed line) and $\Omega_c=2\gamma_0$ (solid line).}
\label{fig_susceptibility}
\end{figure}
The onset of the AT is demonstrated in Fig. \ref{fig_susceptibility}. The system reveals initially a normal Lorentzian peak under $\Omega_c \ll \gamma_1$ but a dip appears as we increase the control laser power $\Omega_c$. Moreover, the width of the transparency window increases significantly  for $\Omega_c \gg \gamma_1$,  and carries a signature of Autler-Townes doublet. We expect that the destructive interference between the excitation pathways is reduced due to a large value of $\gamma_1$. It is important to realize that a change in absorption over a narrow spectral range must be accompanied by a rapid and positive change in refractive index due to which a very low group velocity is produced in AT. Therefore, the group velocity for the acoustic field is given by
\begin{figure}[t!]
\includegraphics[width=0.234\textwidth]{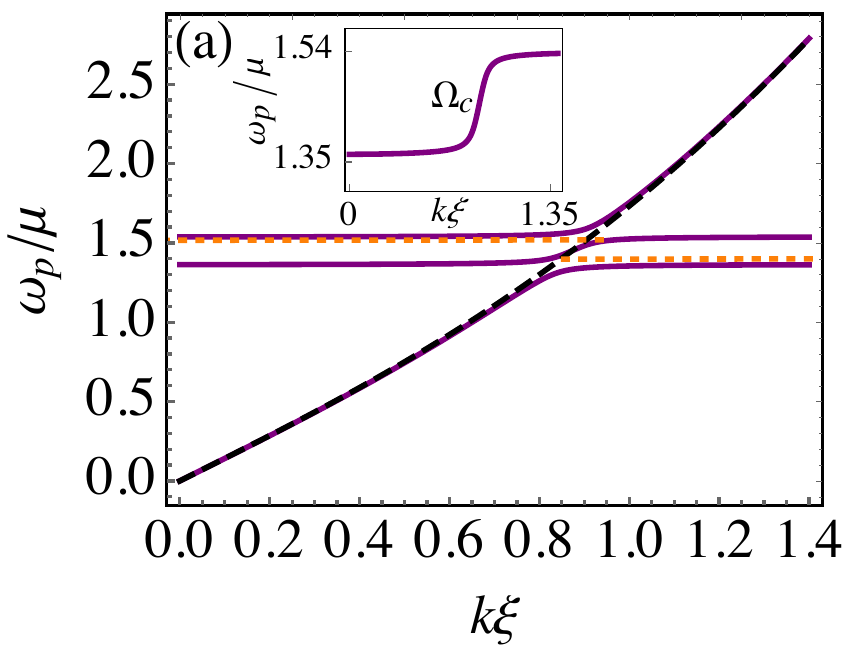}
\includegraphics[width=0.243\textwidth]{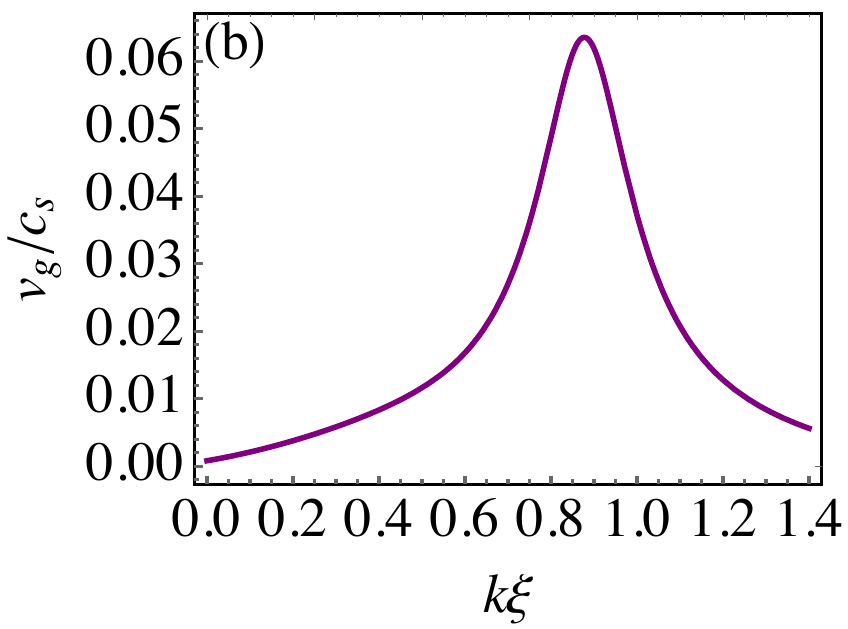}
\caption{(color online) Panel (a) depicts the dispersion curves, where the solid line illustrates the solutions of Eq. (\ref {dispersion}) for $g_{12}=1.85g_{11}$ and dashed line shows the Bogoliubov energy spectrum. The inset shows the dependence of the absorption (transparency) width on the control Rabi frequency $\Omega_c$. Panel (b) determines the group velocity $v_g$  of the order of $0.06$ of the sound speed $c_s$. In all cases, we have set $m_2=1.56m_1$ and a soliton concentration of $N\xi=0.2$.}
\label{fig_DP}
\end{figure}
\begin{eqnarray}
v_g=\frac{c_s}{1+\frac{\chi_{R}}{2}+\frac{\omega_p}{2}(\partial\chi_{R}/\partial\omega_p)},
\end{eqnarray}
where we assume that $\Omega_{c}^{2} \gg 	\Gamma_{p}\Gamma_{c}$.


\subsection{Slow sound in box potentials} 

For the sake of experimental estimates, we consider a one-dimensional BEC loaded in a large box potential. In a typical trap of size $\sim 100$ $\mu$m, healing length $\xi \sim 0.7$ $\mu$m and sound speed $c_s \sim 1$ mm/s \cite{Gaunt}, we can imagine placing up to 20 well-separated ($d \sim 3.5$ $\mu$m) solitons. Under these conditions, the envelope group velocity can be brought down to a record value of $\sim 0.06$ mm/s, corresponding to the peak appearing in Fig. \ref{fig_DP}. Indeed, for a wavelength compared to a intersoliton separation $d$, the estimated group velocity is $v_g \simeq 5.0$ $\mu$m/s. This is much smaller than that obtained in band-gap arrays \cite{robertson2004} and detuned acoustic resonators \cite{Santillan2014}. In the latter, a sound speed of $\sim9.8$m/s is experimentally reported, which makes our scheme able to produce slow pulses by a $\sim 10^{5}$ smaller factor. \par


\section{Conclusion}

In conclusion, we proposed a scheme for the realization of an acoustic transparency phenomenon with dark-solitons qutrits in a quasi-one dimensional Bose-Einsten condensates, in analogy with the well-known phenomenon of electromagnetically induced transparency. The qutrits consist of three-level structures formed by impurities trapped by the dark solitons. We investigate the spontaneous decay rates to analyze the interference effect of the acoustic transparency, due to which a narrow absorption window, depending on the BEC-impurity coupling, can be achieved. We show that an acoustic pulse can be slowed down to a record speed of $5.0$ $\mu$m/s. We believe that the suggested approach opens a promising research avenue in the field of acoustic transport. In general, the present scheme will motivate numerous applications based on the concept of `slow-' or `stopped-sound', such as quantum memories and quantum information processing \cite{Borges2016, Lahad2017}. \par

\begin{acknowledgments} 
This work is supported by the IET under the A.F. Harvey Engineering Research Prize, FCT/MEC through national funds  and by FEDER-PT2020 partnership agreement under the project UID/EEA/50008/2019. The authors also acknowledge the support from Funda\c{c}\~{a}o para a Ci\^{e}ncia e a Tecnologia (FCT-Portugal), namely through the grants No. SFRH/PD/BD/113650/2015 and No. IF/00433/2015. E.V.C. acknowledges partial support from FCT-Portugal through Grant No. UID/CTM/04540/2013.
\end{acknowledgments}


\appendix
\section{Trapping impurities with dark solitons} 
\label{Trapping impurities with dark solitons} 

We consider a dark soliton in a quasi 1D BEC, which in turn is surrounded by a dilute set of impurities  (see Fig. 1 of the main manuscript). The BEC and the impurity particles are described by the wave functions $\psi_1(x,t)$ and $\psi_2(x,t)$, respectively.  At the mean field level, the system is governed by the Gross Pitaevskii and Schrodinger equation, respectively,
\begin{eqnarray}
i\hbar \frac{\partial \psi _{1}}{\partial t}&=&-\frac{\hbar ^{2}}{2m_1}\frac{
\partial^{2} \psi _{1}}{\partial x^{2}}+g_{11}\left\vert \psi _{1}\right\vert ^{2}\psi
_{1}+g_{12}\left\vert \psi _{2}\right\vert ^{2}\psi _{1}, \nonumber \\ 
i\hbar \frac{\partial \psi _{2}}{\partial t}&=&-\frac{\hbar ^{2}}{2m_2}\frac{
\partial^{2} \psi _{2}}{\partial x^{2}}+g_{21}\left\vert \psi_1\right\vert ^{2}\psi _{2},  \label{gps1}
\end{eqnarray}
Here, the discussion is restricted to repulsive interactions ($g_{11}>0$) where the dark solitons are assumed to be not significantly disturbed by the presence of impurities, which we consider to be fermionic in order to avoid condensation at the bottom of the potential and $g_{12}=g_{21}$. To achieve this, the impurity gas is chosen to be sufficiently dilute, i.e. $\vert \psi_1\vert^2 \gg \vert\psi_2\vert^2 $. Such a situation can be produced, for example, taking $^{134}$Cs impurities in a $^{85}$Rb BEC \cite{Michael2015}. Therefore, the impurities can be regarded as free particles that feel the soliton as a potential
\begin{equation}
i\hbar \frac{\partial \psi _{2}}{\partial t}=-\frac{\hbar ^{2}}{2m_2}\frac{%
\partial^{2} \psi _{2}}{\partial x^{2}}+g_{21}\left\vert \psi _{\rm sol}\right\vert
^{2}\psi _{2},  \label{sch. eq. without soliton s}
\end{equation}
where the singular nonlinear solution corresponding to the soliton profile is  $\psi_{\rm sol}(x)=\sqrt{n_{0}}\tanh \left[ x/\xi \right]$. The time-independent version of Eq. (\ref{sch. eq. without soliton s}) reads 
\begin{equation}
(E-g_{21} n_0)\psi _{2}=-\frac{\hbar ^{2}}{2m_2}\frac{%
\partial^{2} \psi _{2}}{\partial x^{2}}-g_{21}n_{0}{\rm sech}^{2}\left( \frac{x}{\xi }\right) \psi _{2},
\label{eq_reflectionless s1}
\end{equation}
To find the analytical solution of Eq. (\ref{eq_reflectionless s1}), the potential is casted in the P\"oschl-Teller form
\begin{equation}
V(x)=-\frac{\hbar ^{2}}{2m_2\xi ^{2}}\nu (\nu +1){\rm sech}^{2}\left( 
\frac{x}{\xi }\right),  
\label{Potent.}
\end{equation}
with $ \nu=\left(-1+\sqrt{1+4g_{21}m_2/g_{11}m_1}\right)/2$. The particular case of $\nu$ being a positive integer belongs to the class of {\it reflectionless} potentials \cite{john07}, for which an incident wave is totally transmitted. For the more general case considered here, the energy spectrum associated to the potential in Eq. (\ref{Potent.}) reads
\begin{equation}
E_{n}^{^{\prime }}=-\frac{\hbar ^{2}}{2m_2\xi ^{2}}(\nu -n)^{2},
\label{energy eigenstates}
\end{equation}
where $n$ is an integer. The number of bound states created by the dark soliton is $n_{\rm bound}=\lfloor \nu+1+\sqrt{\nu(1+\nu)}\rfloor$, where the symbol $\lfloor \cdot\rfloor$ denotes the integer part. As such, the condition for {\it exactly} three bound states (i.e. the condition for the qutrit to exist) is obtained if $\nu$ sits in the range
\begin{eqnarray}
\frac{4}{5}\leq \nu < \frac{9}{7},  \label{condition}
\end{eqnarray}
\begin{figure}[t!]
\center
\includegraphics[width=0.243\textwidth]{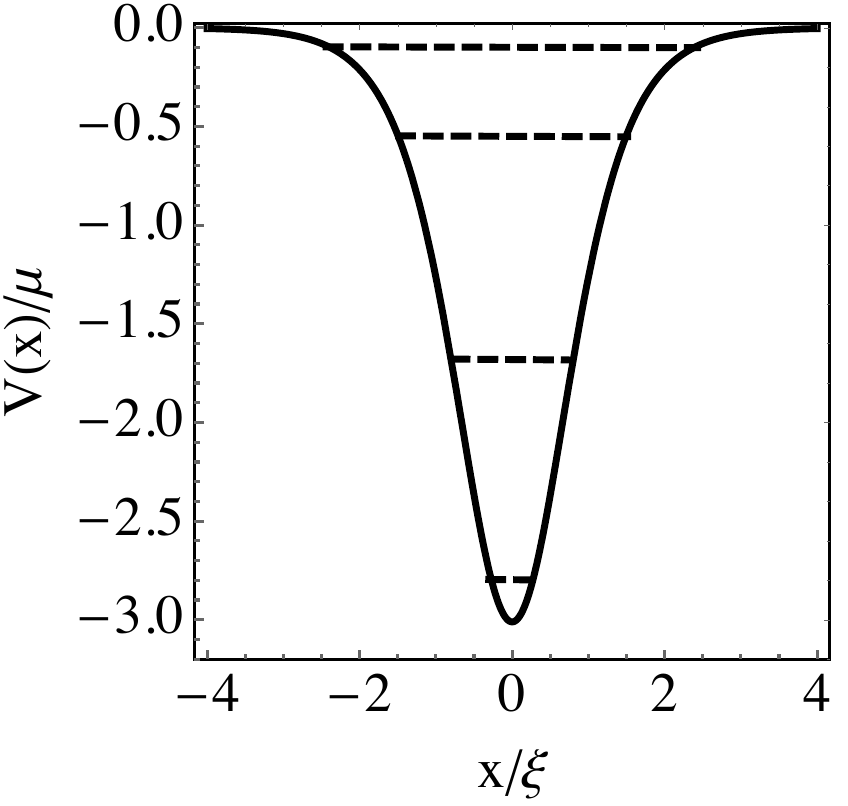}
\caption{(color online) Schematic representation of a four-level system obtained for $\nu=10/7$. The upmost excited state exist at the border of the potential created by the dark soliton.} \par 
\label{states}
\end{figure}
as discussed in the manuscript. At $\nu \geq 9/7$, the number of bound states increases (see Fig. \ref{states} for a schematic illustration). In Fig. \ref{fig_states}, we compare the analytical estimates with the full numerical solution of Eqs. (\ref{gps1}), for both the soliton and the qutrit wavefunctions, under experimentally feasible conditions.  \\\\

\begin{figure*}[t!]
\centering
\includegraphics[scale=0.7]{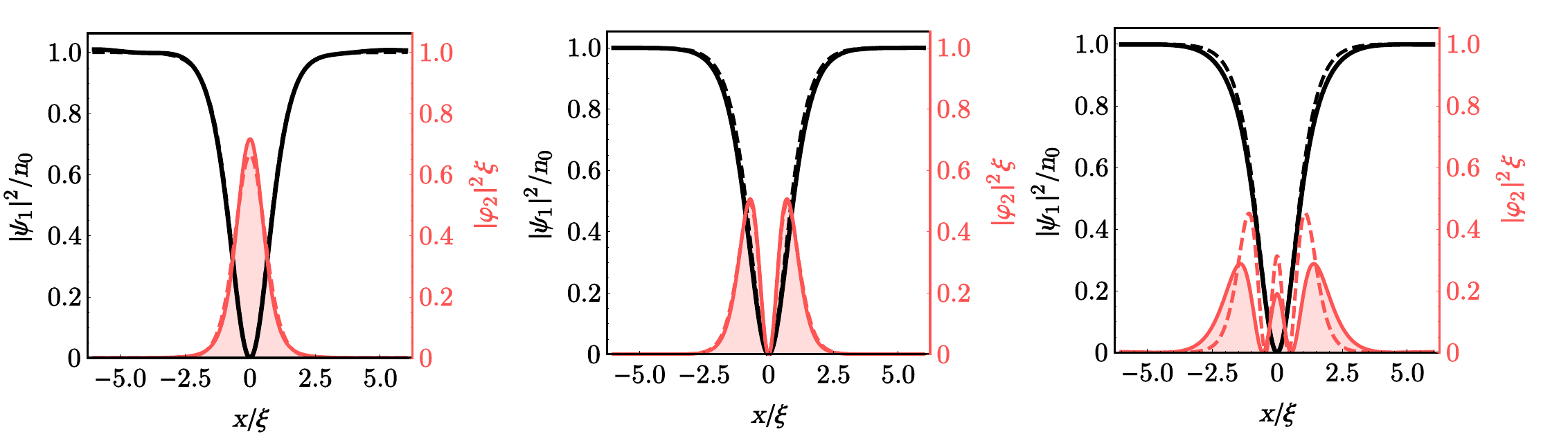}
\caption{(color online) Qutrits in a possible experimental situation: Numerical profiles of the dark soliton (black lines) and the impurity eigenstates (red lines). From left to right, we depict the ground state $\varphi_0(x)$, and the first and the second states, respectively $\varphi_1(x)$ and $\varphi_2(x)$, of a fermionic $^{134}$Cs impurity trapped in a $^{85}$Rb BEC dark soliton. The solid lines are the numerical solutions, while the dashed lines are the analytical expression described in the main text. We have used the following parameters: $m_2=1.56 m_1$, $g_{12}=1.25 g_{11}$ (corresponding to $\nu=1.13$). We fix the number of depleted condensate atoms by the dark soliton to be $n_0\xi\simeq 50$. } 
\label{fig_states}
\end{figure*}

\section{Soliton-phonon Hamiltonian} 
\label{Soliton-phonon Hamiltonian} 

The interaction Hamiltonian is given by
\begin{equation}
H_{\rm int}=g_{12}\int dx \psi_{2} ^{\dag }\psi_1 ^{\dag }\psi_1 \psi_2,
\label{Int. Ham.}
\end{equation}
where $\psi_2(x)=\sum_{l=0}^2 \varphi_{l}(x) a_{l}$ describes the qutrit field in terms of the bosonic operators $a_{n}$, with $\varphi_0(x)=A_0{\rm sech}  ^{\alpha}(x/\xi)$, $\varphi_1(x)=2 A_1\tanh  (x /\xi) \varphi_0(x)$ and $ \varphi_2(x)= \sqrt{2}A_2\left(1-(1+3\alpha)\tanh^{2} (x/\xi) \right)\varphi_0(x)$, where $A_j (j=0,1,2)$ are the normalization constants, given by
\begin{widetext}
\begin{eqnarray}
A_0&=&\left(\frac{\sqrt{\pi}  \Gamma[\alpha]}{\Gamma[\frac{1+2\alpha}{2}]}\right)^{-\frac{1}{2}}, \nonumber \\
A_1&=&\left(2^{2(1+\alpha)}A_0^2 \left(  \frac{ {_{2}}F_1[\alpha,2(1+\alpha),1+\alpha,-1]}{\alpha} 
-\frac{ {_{2}}F_1[1+\alpha,2(1+\alpha),2+\alpha,-1]}{1+\alpha}\right.\right.  \nonumber \\   && \left. \left. 
+\frac{ {_{2}}F_1[2+\alpha,2(1+\alpha),3+\alpha,-1]}{2+\alpha}\right)\right)^{-\frac{1}{2}}.  \nonumber \\
A_2&=&\left(2A_0^2A_1^2\left(   \frac{9\alpha}{2(1+\alpha)}+\frac{9\alpha^2}{4(1+\alpha)} +\frac{9\alpha^2 \sqrt{\pi}(6+5\alpha+\alpha^2)\Gamma[\alpha]}{16\Gamma[\frac{5}{2}+\alpha]}  \right.\right.  \nonumber \\
 && \left. \left. +\frac{3 \times 2^{2(1+\alpha)}\alpha(2+3\alpha) {_{2}}F_1[1+\alpha,2(2+\alpha),2+\alpha,-1]  }{1+\alpha} 
+\frac{4^{(2+\alpha)} {_{2}}F_1[2+\alpha,2(2+\alpha),3+\alpha,-1]  }{2+\alpha} \right.  \right. \nonumber \\
&& \left.\left.+\frac{3 \times 2^{2(2+\alpha)}\alpha {_{2}}F_1[2+\alpha,2(2+\alpha),3+\alpha,-1]  }{2+\alpha}
+\frac{27 \times 4^{(1+\alpha)}\alpha^2 {_{2}}F_1[2+\alpha,2(2+\alpha),3+\alpha,-1]  }{2(2+\alpha)}\right. \right. \nonumber \\
&& \left.\left.+\frac{3 \times 2^{(3+2\alpha)}\alpha {_{2}}F_1[3+\alpha,2(2+\alpha),4+\alpha,-1]  }{3+\alpha}
+\frac{9 \times 2^{2(1+\alpha)}\alpha^2 {_{2}}F_1[3+\alpha,2(2+\alpha),4+\alpha,-1]  }{3+\alpha}  \right. \right. \nonumber \\
&& \left.  \left. +\frac{9 \times 2^{2\alpha}\alpha^2 {_{2}}F_1[4+\alpha,2(2+\alpha),5+\alpha,-1]  }{4+\alpha}
 \right)  \right)^{-\frac{1}{2}},
\end{eqnarray}
\end{widetext}
where ${_{2}}F_1$ and $\Gamma[\alpha]$ represents the Hypergeometic and Gamma function, respectively, and $\alpha=\sqrt{2g_{12}/g_{11}}$. The inclusion of quantum fluctuations is performed by writting the BEC field as $\psi_1(x)=\psi_{\rm sol}(x)+\delta \psi(x)$, where $\delta \psi(x)=\sum_k \left(u_k(x) b_k +v^{*}_k(x)b^{\dagger}_k \right)$ and $b_k$ are the bosonic operators verifying the commutation relation $[b_{k},b^{\dagger}_{q}]=\delta_{k,q}$. The amplitudes $u_k(x)$ and $v_k(x)$ satisfy the normalization condition $\vert u_k(x)\vert ^2 -\vert v_k(x)\vert ^2=1$ and are explicitly given by \cite{Dziarmaga04}, 
\begin{eqnarray*}
&&\left. u_{k}(x)=\sqrt{\frac{1}{4\pi \xi }}\frac{\mu }{\epsilon _{k}}%
\right. \times \\
&&\left. \left[ \left( (k\xi )^{2}+\frac{2\epsilon _{k}}{\mu }\right) \left( 
\frac{k\xi }{2}+i\tanh \left( \frac{x}{\xi }\right) \right) +\frac{k\xi }{%
\cosh ^{2}\left( \frac{x}{\xi }\right) }\right] \right. ,
\end{eqnarray*}
and
\begin{eqnarray*}
&&\left. v_{k}(x)=\sqrt{\frac{1}{4\pi \xi }}\frac{\mu }{\epsilon _{k}}%
\right. \times \\
&&\left. \left[ \left( (k\xi )^{2}-\frac{2\epsilon _{k}}{\mu }\right) \left( 
\frac{k\xi }{2}+i\tanh \left( \frac{x}{\xi }\right) \right) +\frac{k\xi }{%
\cosh ^{2}\left( \frac{x}{\xi }\right) }\right] \right. 
\end{eqnarray*}
Using the rotating wave approximation (RWA), the first order perturbed Hamiltonian can be written as
\begin{eqnarray*}
H_{\rm int}^{(1)} &=&\sum_{k}\left(g_{0}^{k}\sigma^{+} _{0} +g_{1}^{k}\sigma^{+} _{1}\right) b_{k}+ \left( g_{0}^{k*}\sigma^{-} _{0} + g_{1}^{k*}\sigma^{-} _{1}\right)b_{k}^{\dag }, \label{hamilton}
\end{eqnarray*}
where  $\sigma^{+} _{0,1}=a^{\dagger}_{e_1,e_2} a_{g,e_1} $,  $\sigma^{-} _{0,1}=a^{\dagger}_{g,e_1} a_{e_1,e_2} $ and the coupling constants $g_{ll'}^{k}=g_{i}^{k}$($i=0,1$) are explicitly given by

\begin{eqnarray*}
 g_{0}^{k} &=&\frac{i g_{12} k^{2}\xi ^{3/2}}{80\epsilon _{k}}\sqrt{\frac{ n_{0}\pi }{6}}(2\mu +8k^{2}\mu \xi ^{2}+15\epsilon _{k})\times \\ 
 &\times& \left( -4+k^{2}\xi ^{2}\right) {\rm csch} \left( \frac{k\pi \xi }{2}\right), \nonumber\\
g_{1}^{k} &=& \frac{i g_{12} k^{2} \xi^{3/2}}{896 \epsilon _{k}} \sqrt{\frac{n_{0}\pi}{15}}\left[28\left(2k^{4}\xi^{4}-35k^{2}\xi^{2}+68\right)\epsilon _{k} \right.  \nonumber \\
 &+& \left. \mu \left(29k^{6}\xi^{6}-504k^{4}\xi^{4}+896k^{2}\xi^{2}+64\right) \right]    {\rm csch} \left( \frac{k\pi \xi }{2}\right). \label{coupling constants}
\end{eqnarray*}

\begin{figure}[t!]
\includegraphics[width=0.238\textwidth]{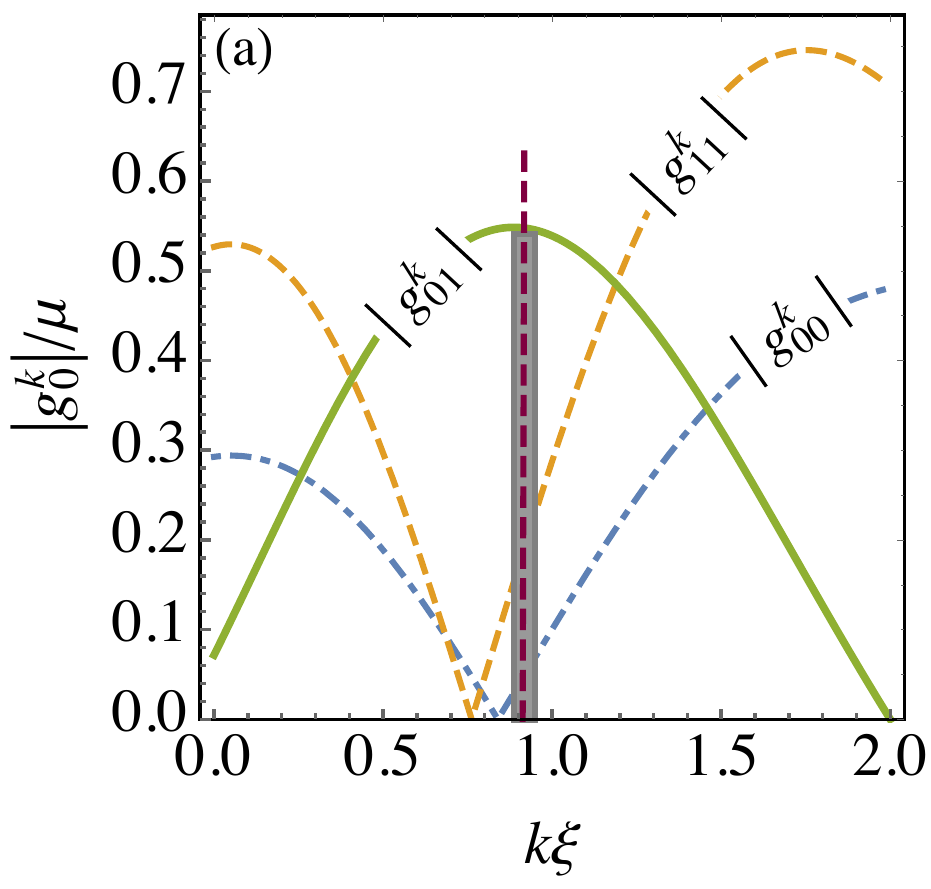}
\includegraphics[width=0.238\textwidth]{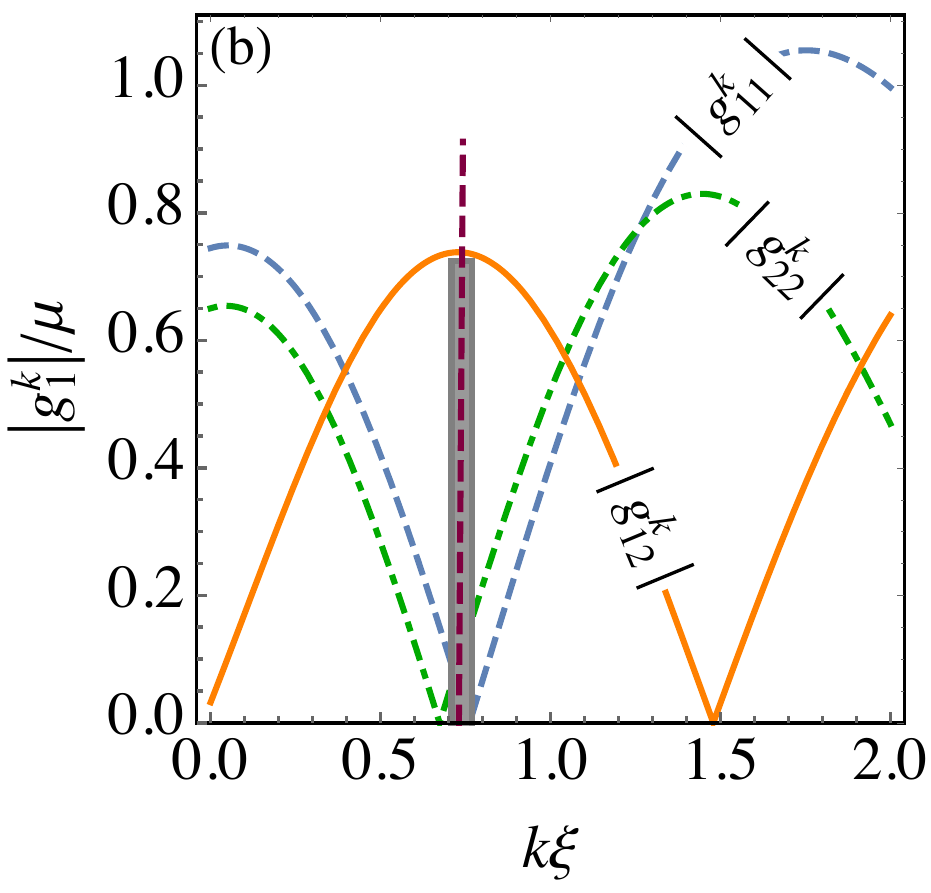}
\caption{(color online) Interband $g_{ll'}^k$ ($l \neq l'$) (solid lines) and intraband $g_{ll'}^k$ ($l = l'$) (dashed or dotted-dashed lines) coupling amplitudes. Near resonance, ($k\sim 0.9 \xi^{-1}$ for the first transition, and $k\sim 0.7 \xi^{-1}$ for the second transition), the interband terms clearly dominate over the intraband transitions, allowing us to neglect the latter within the rotating wave approximation.}
\label{fig_coupling}
\end{figure}
Technically speaking, the RWA approximation here means neglecting the intraband terms in Eq. (\ref{hamilton}), whose amplitudes are given by the coefficients $g_{ll}^k$, illustrated in Fig. \ref{fig_coupling}. This is achieved if we assume that only resonant processes (i.e. phonons with wavectors $k$ such that their energies $\omega_k$ are in resonance with the transitions $\omega_0$ and $\omega_1$, promoting excitation$-$deexcitation of the impurity inside the soliton) participate in the dynamics. As explained in the main text, and as we see below, the validity of our RWA approximation is verified {\it a posteriori}, holding if the corresponding spontaneous emission rates $\gamma_i$ ($i=0,1$) are much smaller than the qutrit transition frequencies $\omega_i$. \\\\

\section{Wigner-Weisskopf theory of spontaneous decay} 
\label{Wigner-Weisskopf theory of spontaneous decay}

We  employ  the  Wigner-Weisskopf  theory to find the spontaneous decay rate of the states, by neglecting  the  effect  of  temperature  and  other  external perturbations. This is extremely well justified in our case, as BECs can nowadays be routinely  produced well below the critical temperature for condensations. The qutrit is assumed to be initially at the excited state $\left\vert e_{2}\right\rangle$ and the phonons to be in the vacuum state $\left\vert 0 \right\rangle$. Under such conditions, the wave function of total system (qutrit + phonons) can be described as
\begin{equation}
\begin{array}{ccc}
\left\vert \phi(t) \right\rangle &=& a(t) \left\vert e_{2},0 \right\rangle + \sum_{k} b_{k}(t) \left\vert e_{1},1_{k} \right\rangle
	\\ &+& \sum_{k,p} b_{k,p}(t) \left\vert g,1_{k},1_{p} \right\rangle, 
\end{array}
\label{qut_state}
\end{equation}
where $a(t)$ is the probability amplitude of the excited state $\left\vert e_{2}\right\rangle$. The qutrit decays to the state $\left\vert e_{1}\right\rangle$ with probability amplitude $b_{k}(t)$ by emitting a phonon of wavevector $k$ and frequency $\omega_k$. Subsequently, the qutrit de-excites to the ground state $\left\vert g \right\rangle$ via the emission of $q-$phonon of frequency $\omega_p$ and probability amplitude $b_{k,p}(t) $. The Wigner-Weisskopf ansatz (\ref{qut_state})  is then let to evolve under the total Hamiltonian in Eq. (\ref{hamilton}), for which the corresponding Schr\"odinger equation yields 
\begin{eqnarray}
\dot a(t) &=&-\frac{\gamma_1}{2 }a(t),  \nonumber \\
\dot b_{k}(t) &=&-\frac{i}{\hbar }g_1^{k\ast }e^{i(\omega _{k}-\omega _{1})t-\frac{\gamma_1}{2}t}-\frac{\gamma_0}{2 }
b_{k}(t), \nonumber \\
\dot b_{k,p}(t) &=&-\frac{i}{\hbar }g_0^{p\ast }b_k(t)e^{i(\omega _{p}-\omega _{0})t^{}},
\label{coefficient soln}
\end{eqnarray}
which are simplified by following the procedure of Ref. \cite{Muzzamal2017}. Here, $\gamma_{i}$ $(i=0,1)$ is the $i$th state decay rate given by
\begin{eqnarray}
\gamma_i &=&\frac{L}{\sqrt{2}\hbar \xi}\int ~ d\omega_k  \frac{\sqrt{1+ \eta_{i}}}{\eta_{i}}\vert g_i^k\vert^2 \delta(\omega_k-\omega_i) \label{eq_gamma},
\end{eqnarray}
with
\begin{widetext}
\begin{eqnarray*} 
\gamma_{0}&=&\frac{\pi N_{0}g_{12}^{2}}{76800\hbar \mu ^{5}\xi ^{2}\eta_0 \sqrt{%
\frac{\mu +\eta_0 }{\mu }}}\left( -\mu +\eta_0 \right) \left( -5\mu +\eta_0
\right) ^{2}
\left( 8\eta_0 +3\mu \left(-2+5\xi \sqrt{\frac{\hbar ^{2}\omega _{0}^{2}}{%
\mu ^{2}\xi ^{2}}}\right) \right) ^{2}\times {\rm csch}^{2}\left( \frac{\pi \sqrt{-\mu
+\eta_0 }}{2\sqrt{\mu }}\right),\\
\gamma_{1} &=& \frac{\pi N_{0}g_{12}^{2}} {2.4\times 10 ^{7} \hbar \mu ^{7}\xi ^{2}\eta_1 \sqrt{\frac{\mu +\eta_1 }{\mu }}} \left( -\mu +\eta_1 \right) \left[-1956\mu^{3}+\hbar ^{2}\omega _{1}^{2} \left[-591\mu+56 \sqrt{\eta_{1}^{2}-\mu^{2}}+29 \eta_{1} \right] \right. \nonumber \\
&+&  \left. 4\mu^{2}\left[505\eta_{1}+7 \sqrt{\frac{\eta_{1}^{2}}{\mu^{2}}-1} \left(107\mu-39\eta_{1}\right)\right]\right]^{2} {\rm csch}^{2}\left( \frac{\pi \sqrt{-\mu +\eta_1 }}{2\sqrt{\mu }}\right), 
\label{eq_gamma0}
\end{eqnarray*}
\end{widetext}
where $\eta_{i}=\sqrt{\mu^{2}+\hbar ^{2}\omega _{i}^{2}}$. In the long time limit $ t \gg \gamma_i$, Eq. (\ref{coefficient soln}) can be simplified to obtain
\begin{eqnarray}
a(t)&=& e^{-\gamma_{1}t/2}, \nonumber \\
b_{k}(t)&=&-ig_{0}^{k}\frac{\left[e^{i(\omega_{k}-\omega_{1})t-\gamma_{1}t/2}-e^{-\gamma_{0}t/2}\right]}{i(\omega_{k}-\omega_{1})-\frac{\gamma_{1}-\gamma_{0}}{2}}, \nonumber \\
b_{k,p}(t)&=&\frac{g_{0}^{k}g_{1}^{k}}{i(\omega_{k}-\omega_{1})-\frac{\gamma_{1}-\gamma_{0}}{2}}
\left[ \frac{e^{i(\omega_{p}-\omega_{0})t-\gamma_{0}t/2}- 1 }{i(\omega_{p}-\omega_{0})-\frac{\gamma_{0}}{2}}  \right. \nonumber \\
&+&\left.  \frac{1-e^{i(\omega_{k}+\omega_{p}-\omega_{eg})t-\gamma_{1}t/2}}{i(\omega_{k}+\omega_{p}-\omega_{eg})-\frac{\gamma_{1}}{2}} \right],  \label{elements s1}
\end{eqnarray}
where $\omega_{eg}=\omega_{0}+\omega_{1}$, leading to Eq. (7) of the manuscript.\\\\

\section{Heisenberg's equation and sound propagation} 
\label{Heisenberg's equation and sound propagation}

With the aim of studying a dilute array of qutrits affects the propagation of sound waves inside the condensate, we compute the equation of motion for a weak acoutic probe coupling the ground and the first excited state (i.e. driving the lower transition of the qutrits). This is done with the help of Heisenberg's relation 
\begin{eqnarray}
i\hbar\frac{\partial \left(\delta \Psi\right)}{\partial t}=\left[\hat{H},\delta\Psi\right], \label{Hies. eq.}
\end{eqnarray}
where $\hat H=H_q+H_p+H_{\rm drive}$ denotes the total Hamiltonian and  $\delta\Psi= \phi b_q e^{iqx}+\psi^* b_q^{\dagger} e^{-iqx}$ is the fluctuating field with the Bogoliubov coefficients $\phi$ and $\psi$. Noticing that the commutation relation with the driving Hamiltonian provides 
\begin{eqnarray}
[H_{\rm drive},\delta\Psi]&=&\frac{g^{k_{res}}_{0}}{2\hbar}\rho_{12},
\end{eqnarray}
where $\Omega_p=N\xi g_{0}^{k_{res}}  \vert \delta \Psi\vert/\hbar$, and proceeding similarly for the commutation with $H_q$ and $H_p$, we obtain the following wave equation 
\begin{eqnarray}
\frac{\partial \Omega_p}{\partial t}+ \frac{\omega_{q}}{q}\frac{\partial \Omega_p}{\partial x}=-\frac{i}{2\hbar^{2}}( g^{k_{res}}_0)^2\rho_{12},
\label{dispersion s1}
\end{eqnarray}
corresponding to Eq. (12) of the main text. The quantum interference with the second transition, driven by the coupling field of intensity $\Omega_c\gg \Omega_p$, is contained in the cohrence $\rho_{12}$ appearing in the rhs of Eq. (\ref{dispersion s1}). The latter can be identified as the acoustic analogue of a dynamical susceptibility.

%
%


\begin{thebibliography}{99}
\bibitem{note1} The present eigenvalue problem can be easily generalised for the case of gray solitons travelling with speed $v$ by replacing $\psi_{\rm sol}(x)=\sqrt{n_{0}}\left[i\theta + \gamma \tanh \left[ x\gamma/\xi \right]\right]$, where $\theta = v/c_{s}$ and $ \gamma = (1-\theta^{2})^{1/2}$. Because this situation is of little use here, we restrict the discussion to the dark-soliton ($v=0$) case only.
\end{thebibliography}
\end{document}